# Spin torque building blocks

N. Locatelli, V. Cros and J. Grollier[*]

Unité Mixte de Physique CNRS/Thales, 1 Avenue Augustin Fresnel, Campus de l'Ecole Polytechnique, 91767 Palaiseau, France and Université Paris-Sud, 91405 Orsay, France

* corresponding author : julie.grollier@thalesgroup.com

The discovery of the spin torque effect has made magnetic nanodevices realistic candidates for active elements of memory devices and applications. Magnetoresistive effects allow the read-out of increasingly small magnetic bits, and the spin torque provides an efficient tool to manipulate - precisely, rapidly and at low energy cost - the magnetic state, which is in turn the central information medium of spintronic devices. By keeping the same magnetic stack, but by tuning a device's shape and bias conditions, the spin torque can be engineered to build a variety of advanced magnetic nanodevices. Here we show that by assembling these nanodevices as building blocks with different functionalities, novel types of computing architectures can be envisisaged. We focus in particular on recent concepts such as magnonics and spintronic neural networks.

The discovery of giant magnetoresistance (GMR) in 1988 [1] [2] laid the foundations for the field of spintronics. In turn, this revolutionized data storage through the development of the GMR hard drive read head, allowing immense storage capacities. But GMR, and more recently tunnel magnetoresistance [3] (TMR)-based sensors, are passive elements dedicated to the readout of magnetic states in nanostructures. A means to actively manipulate the magnetization of nano-objects was provided by the discovery of spin torque (ST), thus promoting spintronic devices to the rank of active elements. Indeed, this effect, which was predicted in 1996 [4] [5] and first observed around 2000 [6] [7] [8] [9] allows for the efficient manipulation of magnetic configurations without the assistance of external magnetic fields, (not compatible with downscaling) through a simple transfer of angular momentum from spin-polarized carriers to local magnetic moments. Consequently, a class of new devices, based on the combined effects of spin torque for writing and GMR or TMR for reading has emerged. The second generation of magnetic random access memories (MRAMs) based on spin-torque writing, called ST-MRAM, is under industrial development and has the potential to replace current cache memory technologies in the next few years thanks to its speed, density, low power consumption and almost unlimited endurance [10]. In this Progress Article, we will show that spin-torque nanodevices are in fact far from limited to binary memories.

## Spin-torque basics

The typical structure of spin-torque devices is a non-magnetic layer sandwiched between two thin nanomagnets (Fig. 1a). One of the layers has its magnetization fixed ($M_{fixed}$), whereas the second one ($M_{free}$) is free to move. When a current is injected through the magnetic stack, the carriers get spin polarized while passing through the ferromagnets. If the magnetizations $M_{fixed}$ and $M_{free}$ are not collinear, as illustrated in Fig. 1a, then the polarized spins incoming in on the free layer are not aligned with $M_{free}$. However, while passing through the free layer, these spins will align with $M_{free}$ due to the exchange interaction. During this process, the spins associated with the conduction electrons lose their component transverse to $M_{free}$. By conservation of angular momentum, this lost spin component is transferred to the free layer in the form of a torque, which is known as the spin-transfer torque. The spin torque can rotate the magnetization of the free layer towards or away from the fixed layer, depending on the sign of the injected current. As predicted in the pioneering works of Slonczewski [4] and Berger [5], the spin-torque amplitude is proportional to the current density, requiring approximately $10^7$ A.cm$^{-2}$ to switch a magnetization at zero field. A decisive advantage of spin torque is that the smaller the device dimensions are, the lower the current that is needed to manipulate the magnetic state. After a decade of intense research and development, the excellent scalability of spin torque has been recently highlighted with low-current (< 30 µA) spin-torque magnetization switching at room temperature in 20-nm diameter junctions, compatible with 22-nm complementary metal-oxide-semiconductor (CMOS) technology [11].

The general principle of spin-torque nanodevices is depicted in Fig. 1b. A current is injected through the trilayer structure. Under the action of spin torque, magnetization dynamics are generated. This magnetization motion is converted into resistance and voltage variations thanks to the trilayer magnetoresistive effect, GMR or TMR, depending on the stack.

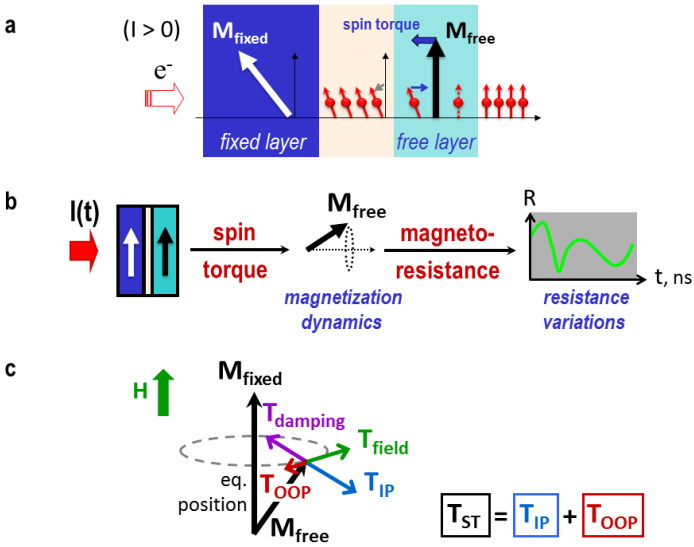

**Figure 1**: **Spin-torque basics. a,** Spin-torque principle: in a ferromagnet /non-magnetic /ferromagnet trilayer, the transverse spin component of the conduction electrons (red) is absorbed as they pass through the free layer, generating a torque on the local magnetization, known as the spin-transfer torque. **b,** Principle of spin-torque nanodevices: when a current is flowing through the trilayer, the spin torque induces magnetization dynamics which are then converted into resistance variations thanks to magnetoresistive effects. **c,** Torques on the local magnetization, under current injection, in the particular case where $M_{fixed}$ and the effective magnetic field are aligned. The two conservative torques, $T_{OOP}$ and $T_{field}$, are aligned, while the two dissipative torques $T_{IP}$ and $T_{damping}$ are parallel. The total spin torque $T_{ST}$ is the sum of $T_{IP}$ and $T_{OOP}$.

The spin torque has two contributions, called in-plane and out-of-plane torques [12] [13], that provide two different handles with which to manipulate the magnetization. As illustrated in Fig. 1c, the in-plane torque $T_{IP}$ lies in the plane defined by $M_{free}$ and $M_{fixed}$, while the out-of-plane torque $T_{OOP}$ points out of it, resulting in very different actions of each torque on the magnetization. The case shown in Fig. 1c, where $M_{fixed}$ and the effective magnetic field are aligned, emphasizes the difference between the torques. In the absence of current, when $M_{free}$ is displaced out of its equilibrium position, it is subjected to the effective-field torque $T_{field}$ that drives it into precession around the effective field, and the damping torque $T_{damping}$, that brings it back to its equilibrium position. When the current is turned on, $T_{IP}$ is aligned with $T_{damping}$ while $T_{OOP}$ is parallel to $T_{field}$. Depending on the current sign, $T_{IP}$ will then either reinforce the damping or act as an anti-damping source. The in-plane torque is therefore useful for stabilizing the magnetization in its equilibrium position, or, on the contrary, to destabilize it to bring it to another equilibrium situation. As for the out-of-plane torque, often called field-like torque, it can emulate the action of a field on $M_{free}$, which means that it can modify the energy landscape seen by the magnetization. The current dependence of $T_{OOP}$ is generally more complex than $T_{IP}$. While $T_{OOP}$ is practically zero in metallic spin-valves, it can reach 40 % of $T_{IP}$ in magnetic tunnel junctions [14].

By adjusting the relative amplitude of the in-plane and out-of-plane torques by tailoring material properties and geometry design, as well as the form of the injected current, the voltage response of spin-torque nanodevices can be largely tuned, allowing the implementation of a great variety of functionalities.

## Magnetization dynamics with the in-plane spin torque

Because $T_{OOP}$ is in general smaller than $T_{IP}$, most spin torque devices are based on $T_{IP}$ only, as an anti-damping source providing a means to destabilize the magnetization without modifying the energy landscape. In this case, because magnetization trajectories are constrained by the field-dependent energy profile, three different scenarios can occur depending on the number of equilibrium positions and their relative stabilities. Figure 2 illustrates the classical case of a free layer magnetization with two equilibrium positions at zero field, parallel (P state) or antiparallel (AP state) to the fixed magnetization. The device response can be tuned by adjusting the amplitude of the applied field with respect to the coercive field $H_c$ required to commute $M_{free}$ between the two stable states.

**Hysteretic switching.** At zero or low external magnetic field ($H < H_c$), both P and AP states are stable (Fig. 2a). By changing the current sign, the in-plane torque will successively destabilize the P and AP states, thus commuting the magnetization back and forth between these two local energy minima. The free layer magnetization switching is associated with large and sharp resistance variations. The hysteresis loop shows that when the current is turned back to zero the two states remain stable. This spin-torque-induced magnetization switching at zero field has found a straightforward application in ST-MRAMs and defined a new class of non-volatile binary memories [15].

**Telegraphic switching.** When the applied field gets close to $H_c$, stochastic switching between P and AP states can occur if the magnetization is destabilized by spin torque in one state, whereas it is barely stable under thermal fluctuations in the second state [16] [17] [18] [19]. By modulating the current amplitude, the spin torque strength and thus the mean time spent in each state can be tuned [20], as shown in Fig. 2b. These adjustable dwell times can potentially be used to encode probabilities, the current amplitude providing a control to adjust the odds. This means that spin torque can also be used to engineer controlled stochastic devices, for instance random-number generators [21].

**Sustained microwave precession.** For external fields larger than $H_c$, only one state remains stable: for example the P state as shown in Fig. 2c. When the current is large enough to destabilize the magnetization from the P state, there is no other local energy minima where the magnetization can stabilize. The magnetization then enters a regime of spin-torque-induced sustained precession [22] [23]. In that case, the magnetization orbit is set by the balance between dissipative ($T_{damping}$ and $T_{IP}$) and conservative torques ($T_{field}$ and $T_{OOP}$).

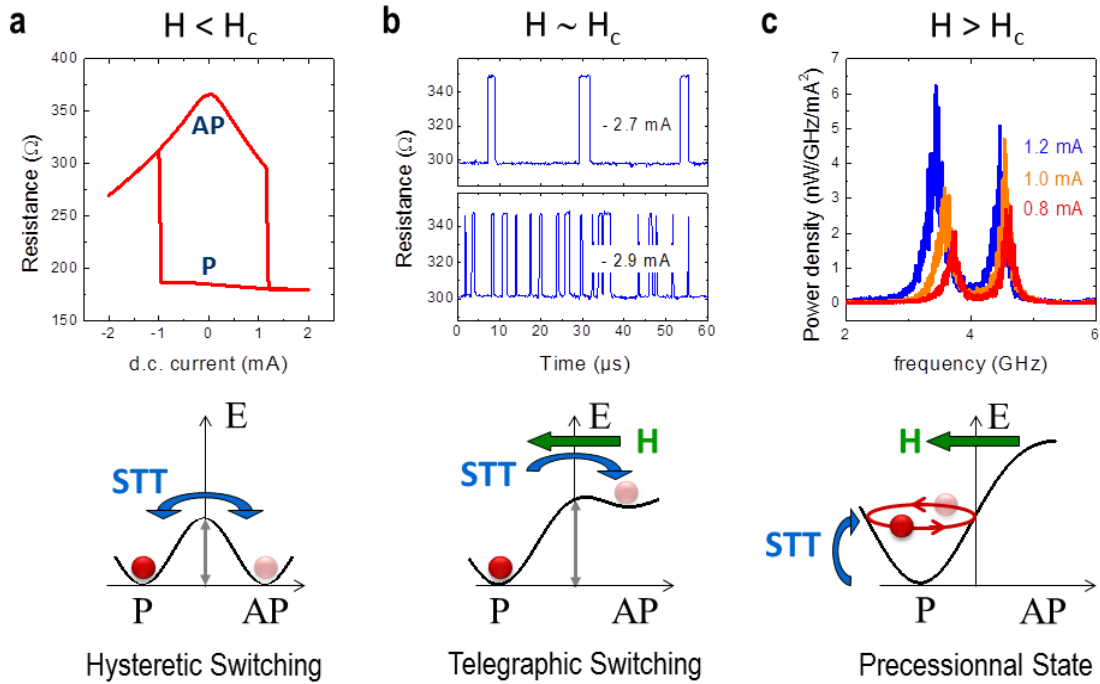

**Figure 2: Magnetization dynamics scenarios under the influence of in-plane spin torque as a function of field amplitude**. Each case is illustrated with experimental results for magnetic tunnel junctions with an MgO barrier and CoFeB electrodes. **a,** When H < $H_c$, spin torque induces hysteretic switching between the two stable equilibrium positions. **b,** For H ~ $H_c$, while spin torque pushes the magnetization out of its most stable position, the external field destabilizes the second equilibrium position, leading to telegraphic switching. **c,** When H > $H_c$ only one stable configuration exists but is destabilized by spin torque: the magnetization is driven into precession on a stable orbit.

## Spin torque bricks

Just as the discovery of GMR boosted data storage in the 1990s, it is envisaged that the sustained microwave precession spin-torque-induced magnetization dynamics can be exploited to build next-generation microwave devices for information and communications technology (ICT). This new class of microwave nanodevices relies on spin torque to induce large-amplitude magnetization precessions and on magnetoresistance to convert these precessions into electrical signals. These spintronic devices have several advantages. First, the free-running frequency, which is linked to the magnetization state and the associated spin-torque-induced vibration mode, depends on the magnetic material and the sample's geometry. By engineering the magnetic systems, a large part of the microwave frequency range can be reached, typically between a few hundred MHz and several tens of GHz. The second advantage is related to their intrinsic nonlinear nature: a simple variation of the injected current will modify the balance between torques, tuning the magnetization orbit and therefore the device frequency extremely rapidly [24] and over a wide range. And finally, the third strength is their deep scalability and robustness to radiations. A unique feature of spin-torque microwave devices is their ability to display multiple functionalities, from signal generation to frequency detection and signal processing. This versatility opens the way for implementing, with the same devices, very different functions such as signl clocks or field sensors in next-generation high-data-rate read-heads [25]. The working principle of each operation is very simple.

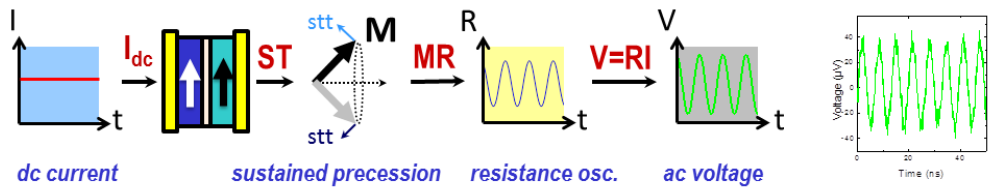

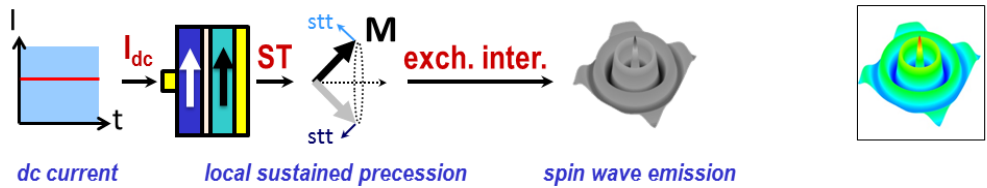

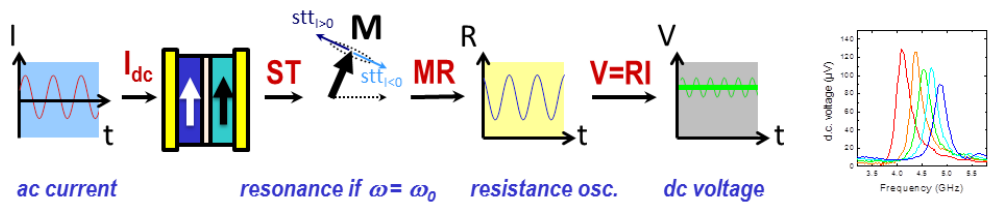

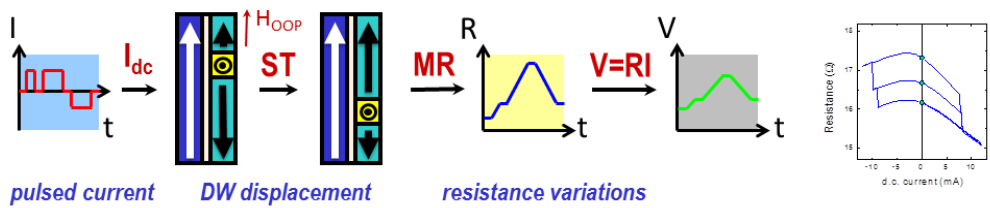

**Figure 3: Underlying principles of spin-torque nanodevices. a,** Spin-torque microwave oscillator: through spin torque (ST), a d.c. current induces sustained oscillations of the magnetization in the free magnetic layer, converted by magnetoresistance (MR) effects in resistance oscillations, and therefore a.c. voltage oscillations. – Right-most panel: experimental time trace of the voltage during current-induced sustained precession of a magnetic vortex in an MgO tunnel junction with a NiFe free layer. **b,** Spin-torque spin-wave emitter: in an extended stack, the d.c. current injected through a nanocontact, via spin torque, induces sustained oscillations of the local magnetization, creating a propagating spin-wave in the extended free layer. – Right-most panel: sketch of a spin wave spatial profile. **c,** Spin-torque microwave detector: through spin torque, an a.c. current with frequency ($\omega$) close to the magnetization resonance frequency ($\omega_0$) induces large-amplitude magnetization precessions. Resistance and current then oscillate at the same frequency, giving rise to a rectified d.c. voltage. – Right-most panel: experimental rectified d.c. voltage versus injected a.c. current frequency for a magnetic tunnel junction with an MgO barrier and CoFeB electrodes, for different amplitudes of the external field (350 to 500 Oe). **d,** Spin-torque memristor: in an elongated shaped trilayer, a pulsed current is used to move a domain wall nucleated in the free magnetic layer through the out-of-plane spin-torque action. The domain wall position modulates the ratio between parallel and antiparallel regions, and therefore tunes the stack resistance, monitored thanks to a small d.c. current. – Right-most panel: experimental resistance versus current hysteretic behaviour of a three-state spin-torque memristor based on an MgO magnetic tunnel junction with a CoFeB fixed layer and a NiFe free layer.

**Spin torque nano-oscillators.** Figure 3a illustrates the principle of spin-torque microwave sources: when the d.c. current induces sustained magnetization precessions, an alternating voltage builds up across the junction. This microwave emission was first demonstrated in magnetic spin valves in 2003 [26] [27]. The transition to large TMR MgO-based magnetic tunnel junctions (~ 100%) in 2008 has allowed an increase of the emitted power above the microwatt range, whereas the first GMR-based devices peaked at a few hundreds picowatts [28] [29] [30]. Such spin-torque nano-oscillators are CMOS compatible, highly integrable, tunable, agile and can even operate at zero field. Furthermore, thanks to their intrinsic nonlinear nature, large bandwidths of frequency modulation can be achieved, which is crucial for signal processing [31]. Spin-torque nano-oscillators are therefore potentially disruptive for telecommunication technologies, but are still the subject of intense academic research because their spectral purity has to be improved. Strategies for decreasing the amplitude and phase noises are the design of innovative phase-locked loops [32] or, at the device level, the dynamic coupling of several magnetic systems. In this trend, it has recently been shown that spin-transfer-driven coupled vortex dynamics can give rise to emission linewidths below 50 kHz at room temperature [33], very close to the upper limit for applications (~ 10 kHz).

**Spin-wave emitters.** Figure 3a illustrates the most widespread oscillator implementation, where the free-layer dimensions are laterally reduced to obtain large current densities, and the oscillating magnetization is confined. An interesting alternative geometry, illustrated in Fig. 3b, is the point contact on an extended free layer, which allows spin-torque-driven emission and propagation of spin waves outwards from the nanocontact [6] [34] [35]. Point-contact spin-torque oscillators are therefore tiny spin-wave emitters.

**Spin-torque microwave nanodetectors.** Signal frequency detection can be achieved by replacing the injected d.c. current with an injected microwave current. During half a period, when the alternating current is positive, $\mathbf{M_{free}}$ is attracted towards $\mathbf{M_{fixed}}$, whereas it is repelled from $\mathbf{M_{fixed}}$ during the second, negative current, half period. As illustrated in Fig. 3c, if the frequency of the injected microwave current closely matches the eigenfrequency of the free layer vibration mode, the induced magnetization motion can be strongly amplified through resonance. The out-of-plane spin torque can also contribute to these dynamics by emulating an alternating microwave field. During this process, the alternating injected current induces resistance oscillations at the same frequency, leading to the appearance of a d.c. voltage signal. This rectification effect called spin-torque diode has been experimentally demonstrated in 2005 and can be used to implement spin-torque microwave nanodetectors [36]. Easily measurable d.c. voltage amplitudes of several hundred microvolts have indeed been reported [37]. The conversion efficiency of the injected microwave power into d.c. voltage can overcome 500 mV/mW, outperforming semiconductor Schottky diodes.

**Spin torque memristor.** After more than a decade of intense research, the understanding of spin torque's microscopic origins and of the resulting magnetization dynamics has reached a level of maturity that now permits an accurate prediction of the device behaviour through coupled transport and micromagnetic simulations. By tailoring the material properties and sample geometries, new bricks can be engineered to obtain a complete set of novel spin-torque-based functionalities. In this regard, the spin-torque memristor is a textbook example of spin-torque device design.

A memristor (short for memory resistor) is a circuit component defined through the expression $V = M(q).I$ relating voltage to current. The "memristance" M is a function of the charge q flowing through the memristor [38]. Memristors are in practice tunable nano-resistors with a non-volatile memory effect. These devices have a strong potential as multilevel digital memories but also as nano-synapses in large scale neuromorphic circuits for fast, low power and defect-tolerant computing [39]. Among the variety of physical effects that have been recently proposed to induce the resistance variations of memristor devices, most are based on deep changes of the internal structure under the application of a voltage (ionic motion, thermal effects and so on) [39] [40] [41]. Alternatively, the spin-torque binary memory presented in Fig. 2a, where the resistance variations are due to reversible magnetic effects, can be seen as a two-level memristor with the associated advantages of speed and reliability.

One possible strategy to realize a multilevel spin-torque memristor is to fabricate a device with an elongated shape to stabilize a magnetic domain wall in the free layer, as illustrated in Fig. 3d. The proportion of P and AP domains can then be varied by displacing the domain wall [42], which will in turn tune the device resistance. Again, spin torque can be used to reliably manipulate the domain wall position, simply by injecting a current perpendicularly to the stack. In this configuration, the in-plane spin-torque action integrated over the domain wall has the symmetry of a field pointing perpendicularly to the stack and is inefficient in moving the domain wall along the wire. The integrated out-of-plane torque action, which instead has the symmetry of a field pointing along the wire, is then essential to move the domain wall [43]. A large out-of-plane torque with a quasi-linear dependence on the injected current can be produced in magnetic tunnel junctions with asymmetric magnetic electrodes [44] [45], allowing the resistance to be increased or decreased at will, by controllably moving the domain wall to the left or to the right, depending on the current sign.

A first proof of concept, given in Ref. [46] and presented in Fig. 3d, shows how the current can allow an efficient resistance control of a three-state spin-torque memristor. The resistance variations are today limited to TMR ratios of about 100 %, but the theoretical limit for $R_{AP}/R_P$ is much higher, close to 300 [47]. The current densities required to move the domain wall up to 500 m/s [48] are a few $10^6$ A.cm$^{-2}$, and it should be noted that critical currents can be reduced below 100 µA by shrinking the device to ~ 20×100 nm$^2$. This crucial step in scalability can be achieved by using perpendicularly magnetized materials with reduced domain wall widths [49].

## Improving Spin Torque devices

At present there is a sustained research effort aimed at improving the characteristics of spin-torque devices. One of the main objectives is the reduction of the currents required to manipulate the magnetization. This is important for decreasing the energy consumption and, if applicable, for shrinking the operating transistors. A first strategy is based on the development and optimization of dedicated materials. Spin-torque devices based on ultra-low damping metallic materials would automatically benefit from a reduction of the in-plane spin torque needed to destabilize the magnetic configuration. Heusler alloys seem to be good candidates for this purpose [50], and encouraging results have already been obtained [51] [52]. There is also currently a lot of work on the development and integration in ST-MRAMs of perpendicular magnetic anisotropy materials [10], to reduce the switching currents while maintaining a good thermal stability. Other studies concentrate on the tunnel barrier properties of magnetic tunnel jucntions, either by working on a deeper understanding of the mechanisms at stake in the prevailing MgO tunnel barrier [53] [54], or by proposing promising substitutes, such as graphene tunnel barriers [55] [56]. Complementary to these materials-based approaches, an alternative strategy is to assist spin torque by additional physical phenomena affecting the magnetization stability. Several possibilities are currently considered and evaluated, such as thermally assisted switching [57], electric field effects on magnetization [58] [59] [60] and spin-orbit torques based on Rashba and/or spin Hall effects [61] [62].

## Spin torque computing architectures

Clearly, spin torque offers the possibility of building nanodevices with a wide range of operations: binary memory, stochastic devices, microwave oscillators, spin-wave emittesr, microwave nanodetectors, memristors, and so on. Indeed a single device can even exhibit different functionalities on demand. As shown in Fig. 2, the operation can be switched between three modes simply by tuning the bias conditions: binary memory, stochastic or microwave device. A crucial advantage of spin-torque devices is that all these functionalities can be obtained using the same materials, the exact same stack, simply by changing the bias conditions or device shape. These different devices can be seen as Lego bricks that can be assembled to build novel types of computing architecture. Figure 4a displays the collection of spin-torque bricks that we have just described. As shown above through the example of the spin-torque memristor, this spin-torque Lego set can be expanded at will.

**Spin-torque logic circuits.** Most potential applications uses a single brick/functionality, and comprises a single device or non-interacting arrays of devices. For example, a hard-disk-drive write-head can include a spin-torque nano-oscillator for the purpose of microwave-assisted switching [63], a radar can use a single tunable spin-torque diode for microwave sensing and an ST-RAM stores the information in arrays of binary memories.

However, due to the non-volatility of nanomagnets, it has long been recognized that nanomagnetism and spintronics have significant potential for low-power processing architectures combining Boolean logic and memory. There are two main research fields in this area.

The first concentrates on hybrid circuits composed of CMOS transistors combined with ST-operated magnetic tunnel junctions. These systems are potentially disruptive as fast and low power logic circuits. Indeed the magnetic tunnel junction devices can be embedded directly on top of the CMOS logic plane, allowing very fast memory access (write and read time below 10 ns), reduced delay times inside the circuit, and most of all sparing the huge energy cost arising from moving bits between memory and logic in standard systems. Furthermore, thanks to the non-volatility of the magnetic tunnel junctions the static power dissipation can be dramatically decreased by suppressing the need to periodically refresh the memory. Spin-torque devices therefore seem good candidates to boost nex-generation logic circuits such as field-programmable gate arrays and application-specific integrated circuits. Prototypes and adapted design tools are currently being developed [64] [65] [66].

A second research field consists of developing ultra-low power spin-logic concepts where logic operations are mostly based on the manipulation of spins and where the need to convert back the information to charges is minimized. Most of these concepts, such as domain wall logic [67], all spin logic [68] or nanomagnet logic whether planar [69], vertical [70] or at the atomic-scale level [71], rely on interacting devices arrays. The information, encoded in the magnetization state of nanomagnets, is read through magnetoresistive effects while spin torque is seen as the ideal replacement to field writing [72].

The spin-torque logic circuits that are developed in these frameworks demonstrate the possibility and interest of co-integrating CMOS/spin-torque devices. Nevertheless, they are restricted to the context of classical Boolean logic and exploit only one of the spin torque operation modes: magnetization switching in binary memories.

**Spin-torque Lego.** Recently, a new class of applications has appeared that takes full advantage of the spin-torque building blocks. The goal here is to assemble different bricks and to combine their various functionalities to build novel types of hybrid spintronic/CMOS information processing hardware architectures working at room temperature with low power consumption and high performances. We will now focus on two such promising innovative architectures: spin-torque based magnonics and neuromorphic architectures. These concepts rely on non-Boolean processing of information. As such they avoid competition with sectors where pure CMOS excels, and open the way for novel types of spintronic accelerators dedicated to specific applications in the field of "Recognition, Mining and Synthesis" [73].

## Spin torque magnonics

Whereas photonics deals with light waves, magnonics uses elementary excitations of spin waves (magnons) to perform calculations through spin-wave emission, manipulation and detection at the nanoscale [74] (Fig.4b). The information can be encoded either in the spin wave amplitude, or in its phase. The original concept relies on microstructured antennas for spin-wave emission, magnetic fields for spin-wave manipulation and inductive methods for spin-wave detection.

Magnonic systems present a number of advantages. First, they offer the possibility of multiplexing several spin waves with different frequencies on the same spin-wave bus [75] which is very important for parallel architectures where performance is defined by the degree of interconnection. They can also benefit from the direct interface with non-volatile nanomagnets as memory elements [76]. Furthermore, spin-wave propagation is fast, easily allowing subnanosecond transmission times. The perspectives in terms of miniaturization are excellent, as the spin-wave length, in the deep submicrometre scale, is several orders of magnitude shorter than for electromagnetic waves [76] with frequencies ranging from GHz to THz. Finally, spin waves can encode and carry information without charges in a digital or analog way. Different types of magnonics logic gates have been proposed depending on the encoding quantity, phase or amplitude. They all rely on the possibility of implementing a spin-wave phase shifter [77]. When the information is the spin-wave amplitude, magnonics logic gates are based on interference processes between inputs [78] [79]. When the information is the phase, the principle of operation is spin-wave superposition. A very recent experimental demonstration of a spin-wave majority gate based on phase coding has been achieved [80]

Although spin-wave decay lengths can be rather long in ferrites, up to centimetres in yttrium iron garnet films [81], they remain limited to a few micrometres in materials that can be easily nanostructured, such as permalloy [35]. This restriction is not necessarily an issue because, in today's chips, the typical interdevice distance is shrinking to the deep submicrometre range. Furthermore, certain types of computation, such as cellular nonlinear networks, take advantage of neighbor-to-neighbour interactions and would be extremely well suited to spin-wave architectures [82].

Although magnonics is not a new concept, it has been very recently proposed that spin-torque nanodevices can be used to redesign all the building blocks of magnonics systems [83] [84]. Spin-torque magnonics logic gates: spin torque based spin wave emitters,

**Spin-torque spin-wave emitters.** As already discussed, nanocontact spin-torque oscillators can be used as spin-wave emitters. Their tiny dimensions, intrinsically under 100 nm, provide a clear advantage compared with wide antennas. It has been demonstrated experimentally that these nanocontacts can emit directional propagating waves [34] [35]. However, to perform magnonic-based logic, it is also necessary to achieve coherent emission of multiple spin-wave beams. For this purpose, it has been proposed to synchronize neighboring nanocontact spin-torque oscillators via the interaction of their spin-waves emission [83], which has already been experimentally demonstrated for two oscillators [85] [86]. Another category of spin-wave emitters is based on the dynamics of magnetic solitons (domain wall, magnetic vortex, and so on). Indeed, when these tiny objects change conformation, collide or annihilate [87] [88], for instance through the action of spin torque [89] [90] [91], they emit spin waves. Spin-torque-driven soliton spin-wave bursting opens the path for ultimate downscaling by setting the emitted spin-waves wavelengths close to exchange lengths, typically smaller than 10 nm, rather than lithography-defined dimensions. These two spin-torque-based strategies for spin-wave emission: ST local magnetization excitation by a nanocontact and ST-induced soliton bursting, are illustrated in Fig.4b.

**Spin-torque spin-wave manipulators.** Spin-wave manipulators rely on a local control of the spin-wave phase or amplitude. Here again, spin torque can be used for this purpose. The local control of the spin-wave phase can be achieved through different strategies. One is to phase lock the spin wave propagating in the spin-wave bus to a driving wave emitted by a spin torque nanocontact placed nearby. We propose as an alternative strategy to use spin torque to move a magnetic domain wall in and out of the spin-wave trajectory. It has indeed been shown theoretically that when a spin wave propagates through a domain wall with a width approximately matching the wavelength, a phase shift is induced [92]. Spin torque can also be used to locally control the spin-wave amplitude. As we have seen, the in-plane spin torque can in some conditions act like a damping or anti-damping source, giving or taking away energy from the spin wave [93]. It has been shown experimentally that indeed, depending on the sign of the injected d.c. current, a spin-torque nanocontact can amplify or attenuate the spin-wave amplitude [94]. The latter strategy for spin-wave manipulation is illustrated in Fig.4b.

**Spin-torque spin-wave detectors.** Depending on the information to be decoded, two spin-torque building blocks can be used: simple magnetoresistive detection of the spin wave passing below the sensing element, which will produce a high-frequency time-varying resistance change; or spin-torque diode microwave detection, which will produce d.c. voltage variations when the spin-wave frequency matches the diode frequency (Fig.4b).

# Spin-torque neuromorphic architectures

Our capacity to build smart multifunctional nanodevices has recently revived the interest in hardware neuromorphic circuits. Neuromorphic systems are inspired by the architecture of the brain. The goal is to analyze and abstract the way biology operates to build computing hardware with one or several of the superior assets of the living brain. For example, the brain is extremely efficient at some tasks that are still out of the reach of sequential von Neumann classical computers, such as very fast face recognition with incomplete data. It has a massively parallel architecture with slow, highly interconnected processing elements. This structure contributes to make it fast, defect tolerant, together with low energy consumption. Hardware artificial neural networks are circuits mapped on silicon that aim at reproducing such precious qualities.

The human brain is composed of about $10^{11}$ neurons and $10^{15}$ synapses. Neurons can be seen as processing units, whereas synapses are adaptive interconnections that define the network memory: the synapse capacity to transmit information, also called the synaptic weight, is adjustable (they are "plastic"), which allows learning. The performance of neuromorphic architectures is linked to their interconnection degree (ratio between the number of synapses and the number of neurons). In hardware, this means that interconnections, that is, synapses, need to be as small as possible. However, CMOS implementations of neural networks typically require static random access memoriy (SRAM) banks to store the synapse weights, plus tens of transistors to mimic their plasticity [95].

**Spin-torque memristors as synapses.** The recent demonstration that a single memristor nanodevice can mimic the synapse behaviour has boosted research in very-large-scale hardware neural networks [96]. Indeed, the plastic synaptic transmission is easily implemented through the memristor non-volatile adjustable conductance. For instance, in the case of the previously described spin-torque memristor, domain wall positions are controlled by spin torque to modulate the device conductance and thus emulate an adjustable synapse transmission (Fig. 4c). Moreover, this simple electric control of memristors' conductance allows the emulation of complex bioinspired learning rules such as spike-timing-dependent-plasticity (STDP) [97]. Replacing CMOS synapses by memristor synapses would yield tremendous gains in terms of silicon area and energy consumption.

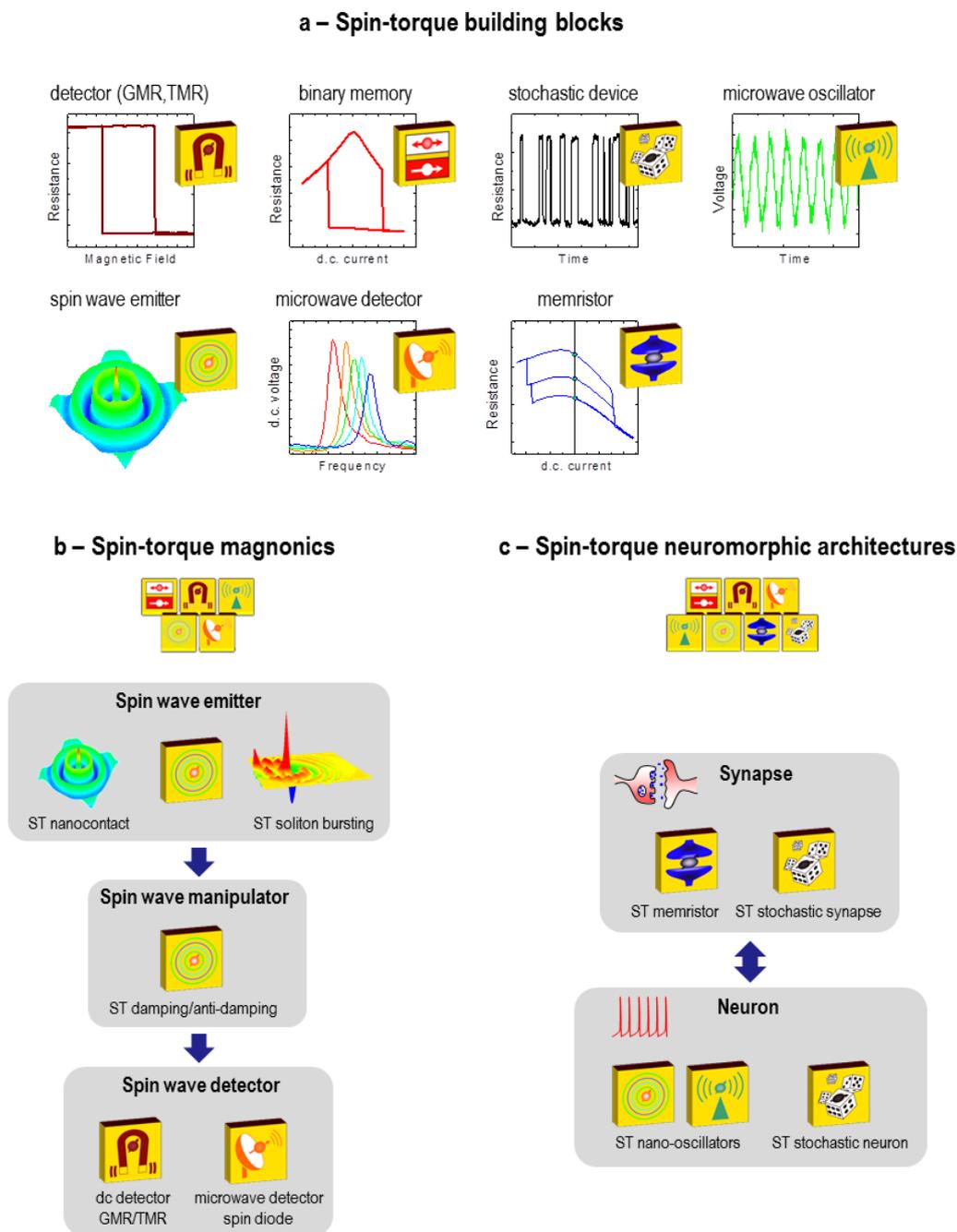

**Figure 4: Spin-torque building blocks. a,** Spin-torque based devices introduced in the text, similar to Lego bricks, with different functionalities that can then be assembled to build novel hardware computing architectures. **b,** Illustration of the working principles of spin-torque magnonics, based on spin-torque induced spin-wave emission, manipulation and detection. Spin-torque spin-wave emission can be obtained through local point contact excitation of a uniformly magnetized extended layer (sketch of a spin-wave spatial profile) or through spin-torque-induced soliton bursting (micromagnetic simulations of the spin-wave emission due to spin-torque-induced vortex core reversal in a magnetic dot). Spin-torque spin-wave manipulation can be achieved by locally damping or anti-damping the spin wave thanks to a spin-torque nanocontact. Spin-torque spin-wave detection can be realized thanks to magnetoresistive or spin-torque diode effects. **c,** Illustration of the main bricks composing spin-torque neuromorphic architectures: the spin-torque synapses and neurons. Different spin-torque-based synapses have been proposed: spin-torque memristors, but also stochastic spin-torque synapses. Similarly, different concepts of spin-torque-based neurons have been recently formulated, exploiting either spin-torque relaxation nano-oscillators or spin-torque-induced stochastic switching.

**Spin-torque nano-oscillators as neurons**. To maximize the neural-networks density, and therefore their efficiency, shrinking neurons to deep submicrometre sizes also becomes an important issue. Recently "neuristor" circuits based on memristors have been used to emulate neurons [98]. Nevertheless these circuits combine several additional standard passive elements including capacitors, known for their large area on silicon. In the following we propose that many neuromorphic building blocks, in particular neurons, can be built efficiently by using single spin-torque-active elements (Fig. 4c). Indeed, the spin-torque memristor is not the only spin-torque brick bridging spintronics and neuromorphic circuits, as there is a clear parallel between neurons and spin-torque nano-oscillators. Biological neurons are spiking cells: when their inputs reach a certain threshold, they emit electrical spikes. They belong to a class of oscillators called relaxation oscillators or "integrate and fire" oscillators [99]. Once again, it has been experimentally demonstrated that under certain conditions, spin-torque oscillators can be converted from harmonic to relaxation oscillators, with features typical of slow charging and fast discharging processes [91]. It should then be possible to engineer an integrate-and-fire spin-torque neuron.

Just like spin-torque oscillators, neurons can be modelled as nonlinear oscillators that adjust their rhythms depending on incoming signals [100]. In the brain, they form a network of coupled oscillators, where the coupling is mediated by synapses. Assemblies of neural oscillators can self-synchronize, in frequency or phase, defining and linking vast areas of the brain where neurons oscillate in unison [101]. Similarly, spin-torque oscillators can self-synchronize when coupled via a mutual electrical [102] [103] [104] or magnetic interaction [85] [86] [105] (demonstrated for up to four oscillators).

**Spin-torque associative memories.** Recent progresses in neuroscience indicate that neural synchronization plays a key role in memory processes [106]. In parallel it has been shown mathematically that neural networks, abstracted as assemblies of synchronized oscillators, can perform associative memory operations on the phase of the oscillators [107] [108]. Models are frequently based on Kuramoto's equation of coupled oscillations, which also describes particularly well arrays of spin-torque oscillators [109]. Associative memories are very different from traditional memories: they are able to retrieve the information on the presentation of noisy or incomplete data. In practice, associative memory processors are important building blocks for applications such as pattern recognition and, in general, classification. These considerations, combined with the tiny size and frequency tunability of spin-torque oscillators, have motivated the proposal of associative memory hardware based on arrays of interacting harmonic spin-torque nano-oscillators [110] [111]. It has been shown that these systems can be decomposed in clusters of reasonable size for the arrays of synchronized spin-torque oscillators without performance degradation [112]. Other authors go even further in the analogy with oscillatory neural networks and propose a general framework for spin-wave-interference-based computation [113].

**Spin-torque neural networks.** Several proposals of full spintronic implementations of neural networks based on nanodevices emulating both neurons and synapses have been recently formulated. Among them, Sharad et al. have developed two different concepts of spintronic neuromorphic hardware [114] [115] [116]; one being an extension of all-spin logic [68] and a second relying on current-induced magnetic domain wall motion. In both cases, the neuron is bipolar and spiking corresponds to the magnetization switching of a magnetic tunnel junction. Krysteczko et al. emulate both synapses and neurons with magnetic tunnel junctions [117]. The synaptic plasticity is achieved by voltage-induced resistive switching phenomena in the MgO tunnel barrier [40], whereas the neuron is emulated by using the magnetic tunnel junction in a regime known as "back-hopping". In that case, the in-plane torque and the out-of-plane torque are opposed, leading to telegraphic switching phenomena similar to the one described in Fig. 2b. The stochastic resistance commutations simulate the neuron bursting behavior.

**Working with stochastic devices.** The controlled stochasticity provided by spin torque is very promising for neuromorphic hardware. Indeed, noise is often seen as a key element of neural computation, beneficial for a number of operations such as near-threshold signaling and decision making [118]. For instance it has been recently demonstrated that spin-torque devices, just like neurons, can exhibit noise-induced sensitivity improvement via stochastic resonance [119] [120]. Stochastic spin-torque elements have a number of additional interesting features.

First they can implement new functions, such as the bursting neuron [117]. Stochastic binary behaviours are also often observed in cell signaling pathways. Spin-torque devices exhibiting telegraph-type behaviour could then also be used to emulate probabilistic biological processes such as neurotransmitter release through the synaptic cleft [121], or the opening/closing of ionic channels [122].

Second, developing processing architectures based on stochastic magnetic devices might allow saving energy. A first strategy is to operate below threshold, that is, to reduce the injected currents below the threshold for deterministic switching. In classical binary memories such as those shown in Fig. 2a, if the injected current is lower than the critical current, the magnetization switching becomes probabilistic and this property can be used to implement probabilistic binary synapses [123]. The strength of synaptic inputs is encoded in the amplitude of the sub-critical current which in turns determines the probability to commute the resistance state. This is another way to express synaptic plasticity, as illustrated in Fig. 4c.

A complementary method to further decrease the energy consumption is to release the demand on the degree of non-volatility. For example, all synapses do not necessarily need to be able of long-term memory. If a network needs a time $t$ to perform calculations, synapses belonging to this network should store the information accurately only during that time $t$. The calculation output alone needs to be stored in a separate long-lasting memory. In terms of spin-torque devices, this means that the energy barrier between the P and AP state can be decreased, resulting in a strong reduction of critical currents for operation.

## Advanced computation based on spin-torque devices

In conclusion, we have emphasized that assembling a few spin torque bricks can be used to build novel types of computing architectures. Here we have provided a few examples such as spin-torque magnonics or spin-torque neuromorphic systems but others remain to be invented. A lot of fundamental investigations will have to be performed before the first hybrid spintronic/CMOS hardware prototypes based on spin torque will actually be developed to design an advanced computation scheme. Nevertheless, fascinating developments can be foreseen owing to the versatility and the scalability of spin-torque effects, notably through the handles they provide to stochastic and even chaotic phenomena. Such non-deterministic behaviours, which will in any case become increasingly important as device dimensions shrink, are indeed not seen as detrimental anymore but allow new spin-torque bricks to be defined for emulating the richness and complexity of neural networks. The spin-torque devices we have outlined here should provide the building blocks to a rich and diverse set of ICT devices and computing architectures.

**Acknowledgements**

We would like to acknowledge the spin torque team at Unité Mixte de Physique CNRS/Thales, especially Albert Fert and all present and past students. We thank all the team of Shinji Yuasa in AIST Tsukuba Japan for invaluable collaboration. We are grateful to Olivier Temam, Damien Querlioz, Pierre Bessière and Jacques Droulez for stimulating discussions. J. Grollier and N. Locatelli acknowledge financial support from the European Research Council (ERC "NanoBrain" 2010 Stg 259068).

**Competing financial interests**

The authors declare no competing financial interests.